\documentclass{egpubl}
\usepackage{eurovis2024}

\SpecialIssuePaper         

\AuthorPreprint

\usepackage[T1]{fontenc}
\usepackage{dfadobe}  

\usepackage{cite}  
\BibtexOrBiblatex
\electronicVersion
\PrintedOrElectronic
\ifpdf \usepackage[pdftex]{graphicx} \pdfcompresslevel=9
\else \usepackage[dvips]{graphicx} \fi

\usepackage{egweblnk}

\usepackage[english]{babel}
\addto\extrasenglish{%
}

\usepackage{multirow}
\usepackage{tabularx}

\title[ChoreoVis]%
      {ChoreoVis: Planning and Assessing\\Formations in Dance Choreographies}

\author[S. Beck, N. Doerr, K. Kurzhals, A. Riedlinger, F. Schmierer, M. Sedlmair \& S. Koch]
{\parbox{\textwidth}{\centering
    Samuel Beck\orcid{0000-0003-0596-6333},
    Nina Doerr\orcid{0000-0003-3249-5354},
    Kuno Kurzhals\orcid{0000-0003-4919-4582},
    Alexander Riedlinger,
    Fabian Schmierer,
    Michael Sedlmair\orcid{0000-0001-7048-9292}
    and Steffen Koch\orcid{0000-0002-8123-8330}
}
        \\
{\parbox{\textwidth}{\centering
         University of Stuttgart, Germany
       }
}
}

\begin{document}

\teaser{
 \includegraphics[width=\linewidth]{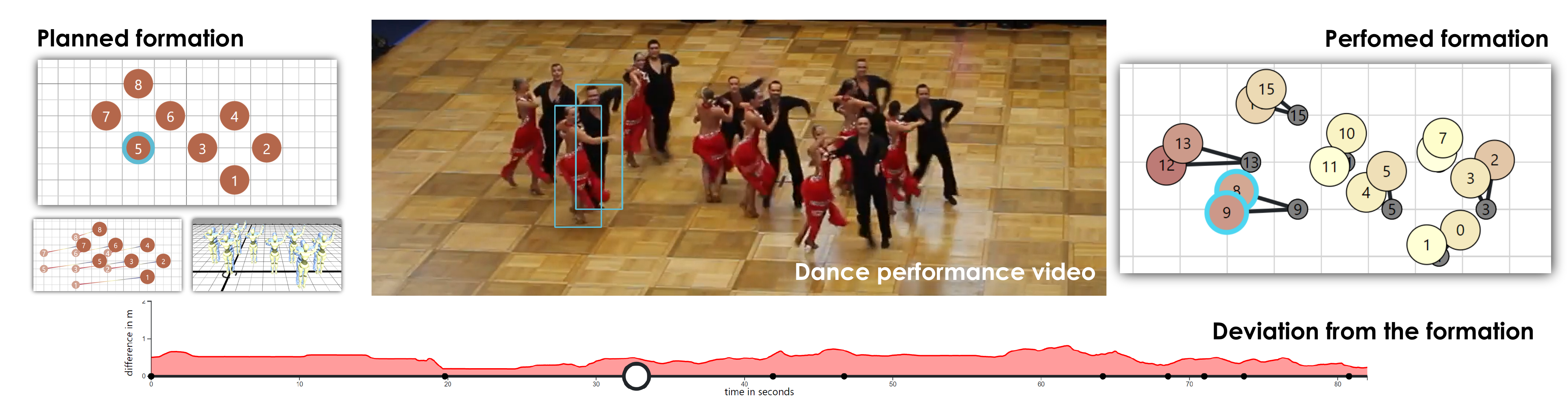}
 \centering
  \caption{Formation in a dance choreography as a planned schematic on the left, in the real performance in the middle, and a visual comparison of the two on the right. The timeline at the bottom gives an overview of the team's precision throughout the performance.}
\label{fig:teaser}
}

\maketitle

\begin{abstract}
Sports visualization has developed into an active research field over the last decades. Many approaches focus on analyzing movement data recorded from unstructured situations, such as soccer.
For the analysis of choreographed activities like formation dancing, however, the goal differs, as dancers follow specific formations in coordinated movement trajectories. To date, little work exists on how visual analytics methods can support such choreographed performances.
To fill this gap, we introduce a new visual approach for planning and assessing dance choreographies.
In terms of planning choreographies, we contribute a web application with interactive authoring tools and views for the dancers' positions and orientations, movement trajectories, poses, dance floor utilization, and movement distances.
For assessing dancers' real-world movement trajectories, extracted by manual bounding box annotations, we developed a timeline showing aggregated trajectory deviations and a dance floor view for detailed trajectory comparison.
Our approach was developed and evaluated in collaboration with dance instructors, showing that introducing visual analytics into this domain promises improvements in training efficiency for the future.


\end{abstract}  

\section{Introduction}\label{sec:introduction}
The representation and analysis of data from different sports have developed into a growing research field in the visualization community~\cite{perin2018, du2021survey}.
It is spread across a multitude of different categories; popular ones include soccer analysis~\cite{stein2018}, table tennis~\cite{wu2017ittvis}, and also categories focusing more on individuals, such as running~\cite{coenen2021public}.
In this work, we focus on choreographed group dancing using the example of Latin formation dancing, a lesser-known team sport that has only seen little visualization-related research so far.
In this sport, formation teams of up to eight couples compete with each other by performing choreographies that consist of the five Latin dances: Cha Cha Cha, Rumba, Samba, Jive, and Paso Doble.
Internationally, the sport is regulated by the \textit{World DanceSport Federation}~\cite{WDSF2023} that includes national regulatory bodies like the \textit{German Dance Sport Association} (DTV)~\cite{DTV1998}.
Dancing coordinated choreographies with multiple couples performing the same dance steps to create a coherent impression presents additional challenges over those already found in single couple dancing.
Important scoring factors are technical dancing ability and difficulty, synchronicity, the formations formed on the dance floor, and the transitions between them.
The latter are a large part of the choreographic performance.
Each dancer or couple has an assigned position on the dance floor to create moving and momentarily stationary formations.
Such formations include lines, diagonals, squares, or diamonds.
According to the DTV~\cite{DTV1998}, they are scored based on the precision of lines, equality of distances, symmetry, variety, and whether they are moving or stationary.
Planning the formations in a dance choreography and the transitions between them, communicating them to the team, practicing them, and assessing them during training and performances are some of the most challenging aspects of formation dancing.
Planning takes place on paper or with the help of rudimentary tools.
Formation definitions are often shared as static PDF files, which are hard to update and make it difficult to analyze the formation transitions before training begins.
The assessment of the precision of formations in performances is only supported through video recordings.
Visual analytics methods can provide means to support such tasks for choreographed formation dancing.

In this work, we expand on earlier research \cite{beck2023visual} to fill this gap and present a design study that introduces a new visualization approach for planning and assessing the formations in dance choreographies.
Although the design study was conducted with the example of Latin formation dancing, the presented approach should be suitable for most group dances that give importance to the spatiotemporal arrangement of dancers.
To gain a deeper understanding of the tasks and requirements in this domain and to design an approach that fulfills the needs of formation teams, we collaborated closely with three formation instructors.
As a result of our work, we make three primary contributions.
First, we present a detailed description of tasks and requirements for planning and assessing formations in the choreographed dancing domain.
Second, we introduce \emph{ChoreoVis}, a visual analytics approach for planning and assessing the formations in group dance choreographies developed in close collaboration with formation teams and instructors.
Finally, we discuss a case study of \emph{ChoreoVis} to demonstrate its applicability and benefits to the iterative workflow of planning, sharing, practicing, and assessing group dance formations.
In addition, we present the results of an expert study of our choreography planning approach that helped us to improve the approach further.

Our collaboration and results show that using visual analytics methods in choreography planning and assessment promises improvements in training efficiency.
The developed prototype for planning was well received.
\emph{ChoreoVis} has been successfully employed by our collaboration partner in the last eight months to plan and train their new formation choreography, with more study participants voicing their interest in using the approach for their training.
We continuously improve \emph{ChoreoVis} with new features and quality enhancements thanks to the ongoing collaboration.

\section{Background and Related Work}\label{sec:relatedwork}

Our work is related to several subareas in visualization research. 
They are \emph{sports visualization} where the depiction and \emph{analysis of movement data} often play an important role. 
Furthermore, in our work, movements are extracted from and \emph{visualized in video} footage. 
Besides that, our approach offers \emph{visual authoring support}.
Finally, since dance choreographies are always planned and aligned to music, \emph{music visualization} that typically uses some sequential \emph{time-based representation} also deserves consideration.

\subsection{Video, Sports, and Movement Visualization}

We present a visualization approach combining abstract visualization for planning and video-based analysis for assessment. Hence, we see our work grounded in the field of video visualization~\cite{afzal2023visualization,borgo2012state} and video visual analytics~\cite{hoferlin2015scalable,tanisaro2015visual}. We address trajectory data extracted from videos, which was investigated in numerous domains (e.g., surveillance~\cite{hoeferlin2013,meghdadi2013interactive}, medicine~\cite{duffy2013glyph}, sports~\cite{pingali2001,stein2018}) but not for the assessment of pair dance performances, which poses multiple new challenges for visualization design, especially with a target group not consisting of visualization experts. 

In their survey about data visualization in sports, Perin et al.~\cite{perin2018} categorized techniques based on (1) the data, (2) the sport, (3) the scientific contribution type, and (4) the included evaluation.
According to this scheme, our approach focuses on tracking data.
Academic publications in this category cover all famous sports from soccer~\cite{stein2018}, tennis~\cite{polk2014tennivis,polk2019}, basketball~\cite{chen2016}, and baseball~\cite{dietrich2014baseball4d}, to cycling~\cite{wood2015visualizing}, and running~\cite{coenen2021public,napolean2019running}. Only one publication is listed as \textit{other} sports, i.e., table tennis~\cite{wu2017ittvis}. 
The body of existing work comprises visual analysis approaches for individual and team behavior. Most of these disciplines have in common that the sports actors react in specific situations and analysis focuses on the detection and interpretation of such events.
With Latin formation dance, we investigate a new sports category that has been barely addressed in visualization. Compared to other sports, event sequences here are strictly planned and have to be coordinated between dancers.
In terms of contribution, we categorize our work by providing a design study that involves sports actors and a new visualization technique covering the most important steps to support dance instructors. The evaluation is categorized by a case study and collaboration with sports actors, which are common methods according to the survey.
Lin et al.~\cite{lin2023ball} additionally stress the importance of closely working with sports experts when designing sports visualizations.

Our planning approach includes views for analyzing athlete performances via heat maps and trajectory maps. Such visualizations are common in sports like basketball~\cite{Losada2016} and soccer~\cite{stein2018} but have yet to be applied to formation dancing.
Moreover, their target group is often spectators rather than sports actors.
However, Page and Moere~\cite{page2006} call for more visualization techniques tailored to athletes.
Menzel et al.~\cite{menzel2023beat} introduce a technique for automated analysis and evaluation of synchronicity in dance performances, an important factor in formation dancing.
Their method is also video-based and enables the automated comparison of extracted body poses with an audio track to identify whether the dancer is on beat.
Another technique that proposes the video-based analysis of dance performances is DanceVis by Guo et al.~\cite{guo2022dancevis}.
Similar to our methodology, they collaborated with experts in designing their technique, which is focused on evaluating a dancer's performance by automatically calculating scores for different criteria.
Neither approach considers dance groups or concerns itself with dance formations.
Instead, they focus on movements and synchronicity of individual dancers with the music.

\subsection{Interactive Authoring, Music and Time Visualization}

Interactive visual authoring is an indispensable tool in numerous domains today, including graphics, photo and video editing, and many more.
Authoring approaches gain more interest in visualization research as well, either as a means to help users create data visualizations~\cite{ren2018charticulator}, convey visual results through storytelling~\cite{tang2020plotthread}, or augment and support the generation of other products.
The degree of freedom in terms of how much the outcome can be influenced by authors strongly varies depending on the difficulty and abstraction of the task, from very fine-grained, manual editing possibilities to strongly machine learning-supported processes with a higher level of automation.
Here, we particularly focus on authoring approaches that consider temporal information and spatial positioning tasks.

Recently, Offwanger et al.~\cite{offenwanger2023timesplines} suggested an authoring tool for temporal information that gives authors more freedom regarding the orientation and placement of visually represented timelines and associated information.
Tang et al.~\cite{tang2020plotthread} proposed a visual interactive system with machine learning assistance for supporting visual storytelling tasks. 
Our approach incorporates an authoring component that supports trainers in editing and planning new choreographies.
Such planning tasks include setting up the positions of couples in dance formations, considering the transitions between such formations, and aligning them to the music and overall sequence of the choreography.
As opposed to many of the mentioned tools for telling stories to broader audiences, the component in our approach focuses on creating an abstract choreography description that supports dancers and instructors. 

Existing research prototypes with a focus on authoring dance choreographies, like ChoreoGraphics from Schulz et al.~\cite{schulz2013} and DanceStudio by Muhammad~\cite{muhammad2009}, allow to capture dance movements, create simple formations based on constraints, and check transitions for collisions.
There are also some commercial tools for choreography authoring like StageKeep~\cite{stagekeep}, FORMI~\cite{formi}, and ArrangeUs~\cite{arrangeus} that offer more polished user experiences, support collaboration, mobile use, and synchronization with music.
They focus on defining the dancers' positions and animating transitions between them.
However, existing tools lack the extensive analysis features and views proposed in our approach and offer no solutions for assessing performances.

The visual representation of temporal data has been a research area in visualization for a while now~\cite{aigner2011visualization, brehmer2016timelines}.
Choreographies are designed and planned to fit the music.
Musical data can be seen as an inherently temporal or sequential type of data. 
Accordingly, this sequence needs to be represented in a respective authoring tool.
Different from other visual representations of musical data~\cite{khulusi2020survey}, the alignment of choreographic formations with the musical part in our approach takes place on an abstract level.
The realization of this sequential aspect and the alignment of the choreography to it is therefore realized in a more abstract way than 
usual when supporting the analysis of sheet music~\cite{miller2022corpusvis}, composing music~\cite{rau2022visualization}, music recommendation~\cite{saito2011musicube}, or practicing musical performances~\cite{heyen2022augmented}.
Our approach uses discrete modeling~\cite{aigner2011visualization} of music (play time) in a linear timeline representation.
Only the bars of the piece, which need to be aligned with the dance formations of the choreography, are shown. 

\begin{figure*}[t]
    \centering
    \includegraphics[width=0.9\linewidth]{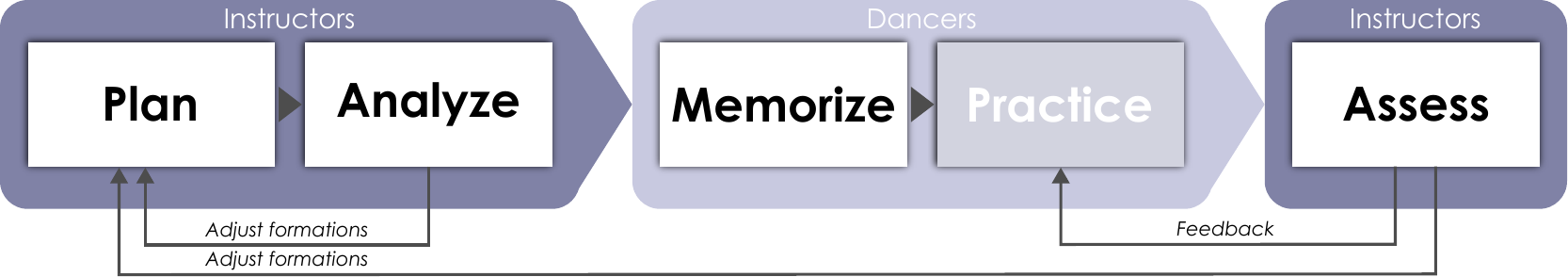}
    \caption{The iterative process of choreographing and practicing dance formations. 
    \emph{ChoreoVis} directly supports all steps except the practice of the choreography. We mainly address instructors, but the planning application can also be utilized for memorizing the formations.}
    \label{fig:pattern-workflow}
\end{figure*}

\section{Domain and Requirements}\label{sec:technique}

In the development of \emph{ChoreoVis}, we followed a user-oriented approach that heavily leaned on the design study methodology proposed by Sedlmair et al.~\cite{sedlmair2012}.
We first conducted group interviews with experts to understand the tasks and problems in this domain, identify design requirements, and determine what data has to be processed.
Consequently, our design resulted in two components: (1) a visualization approach for interactive planning and (2) a video visualization to link the actual performance with the planned choreography and assess the deviations from it.
For the former, we conducted two interviews for requirements elicitation and to collect feedback as well as a think-aloud study for the evaluation.
For the latter, we also conducted a group interview to elicit requirements and collect feedback on mock-ups of our visualization ideas. The transcripts are included in the supplemental material.

\subsection{Method}\label{subsec:method}
In the following, we describe the three semi-structured group interviews~\cite{wood1997semi} that we conducted with up to three experts with 24 years of combined experience as formation instructors.
Transcripts were produced for each interview, which are part of the supplementary material.
Based on these, we derived a workflow for choreographing and practicing dance formations, detailed in \autoref{subsec:tasks}, and the requirements for our approach, listed in \autoref{subsec:requirements}.
In addition, we had frequent informal conversations with instructors and dancers to deepen our understanding of the domain and receive feedback for our approach.
Finally, to evaluate the interactive planning prototype, we conducted a think-aloud study~\cite{boren2000thinking} with four experts that we report on in \autoref{sec:evaluation}.

\paragraph*{Interview 1}
The goal of our first interview was to gather requirements for the visualization approach for interactive planning.
We also gathered general information about the existing process of creating choreographies and formations and about previous experience with digital tools.
As a result, we documented the workflow described in \autoref{fig:pattern-workflow} and current limitations and specified the requirements R1 to R6.

\paragraph*{Interview 2}
During the development phase, we presented an early version of the \emph{ChoreoVis} planning prototype to two of our experts to collect feedback.
During this one-hour session, we mainly received positive feedback from the experts.
Our demonstration also inspired additional feature requests, mainly concerning the usability of the editing features, which led to the refinement of requirement R2.

\paragraph*{Interview 3}
A third interview with two of our experts, which lasted around one hour, aimed to understand their evaluation of the precision of the formations in a performance.
Subsequently, we elicited requirements for a video-based trajectory comparison approach to aid in this workflow and discussed first mock-ups of our \emph{ChoreoVis} assessment prototype to refine the design of the approach.
As a result, we gained insights into the \emph{assess} step of our workflow and specified the requirements R7 to R9.

\subsection{Tasks, Workflow, and Current Limitations}\label{subsec:tasks}

In the following, we describe our understanding of the domain gained through the collaboration and interviews with formation instructors (\autoref{subsec:method}).
We identified a workflow regarding the planning, training, and assessment of formations that is illustrated in \autoref{fig:pattern-workflow}.
It comprises tasks for formation instructors and dancers.
Instructors first \emph{plan} and \emph{analyze} the formations in a choreography. Dancers then \emph{memorize} and \emph{practice} the formations during training.
Finally, the instructors \emph{assess} the dance performance to identify issues that have to be addressed by more practice or by adjusting the formations.

\paragraph*{Planning Formations}
We learned that planning the formations and transitions is usually the second step in creating a dance choreography.
Choreographing the actual dance movements is the first priority.
Once they are set, choreographers and instructors move on to planning formations that fit the movements to create an overall coherent dance impression.
Our domain experts have different approaches to planning the formations in their choreographies.
Some prefer to brainstorm by drawing them in a Cartesian coordinate system on paper first and digitizing them later.
Others directly use a software tool to try out different ideas.
During our first interview, the experts explained that the currently employed tool is proprietary, not publicly available, and no longer maintained.
Its functionality is limited to the definition of the individual positions within formations without the possibility to define further characteristics like orientations, transitions, and point definitions.
Moreover, it lacks our approach's analysis features, and choreographies can only be shared as static PDFs.
Furthermore, planning the formations is often a collaborative process, but the tool's proprietary data format and offline nature hinder successful collaboration between instructors.

\paragraph*{Analyzing Formations}
Features that aid the theoretical analysis of the formations and transitions in this early design stage are mostly missing from currently employed tools.
The consequences are many iterations of the choreographed formations because issues are not discovered before practice.
This time-consuming and frustrating process for the team could be made more efficient by offering analysis techniques to identify and mitigate potential problems before practice.

\paragraph*{Memorizing Formations}
Once the choreography has been planned, the formations are shared with the team so that they can memorize their positions.
Currently, this is done in the form of complex and lengthy PDF files that can have up to hundreds of pages.
For each formation, they usually contain a drawing in a coordinate system, a table listing the positions, and additional remarks from the instructors.
When memorizing their positions, dancers often focus on behind or next to whom they are supposed to stand in a formation in addition to their absolute position.

\paragraph*{Practicing Formations}
To practice the realization of the formations in their dance performance, teams place distance markers around the dance floor as a guide for arranging themselves in the formations.
Dancing short passages and assessing the formations is repeated frequently, making this a critical cycle in the practice workflow.
Furthermore, we noticed it is common for dancers to look up their positions during training sessions.
Therefore, providing responsive solutions that can be accessed from mobile devices during practice is essential.
When a team starts to learn a new choreography, it is common to discover problems in the formation transitions that went unnoticed during the planning stage.
As adjusting the formations is time-consuming, minimizing the necessary changes during practice would lead to a more streamlined and efficient training experience.

\paragraph*{Assessing Formations}
Instructors constantly assess the team's performance during practice sessions to identify mistakes and passages that require more training.
To this end, they use distance indicators around the dance floor to estimate if dancers are standing on their assigned positions.
However, more importantly, they look at the team as a whole, often from an elevated viewpoint, to spot formations that are off.
A common tool to aid the assessment is recording and analyzing videos of the performance to identify issues.
Videos also offer the possibility for dancers to assess their own performance and to spot opportunities for improvement.
Beyond that, our experts were unaware of any existing software tools that support assessing choreographed dance performances.

\begin{figure*}[t]
    \centering
    \includegraphics[width=0.9\linewidth]{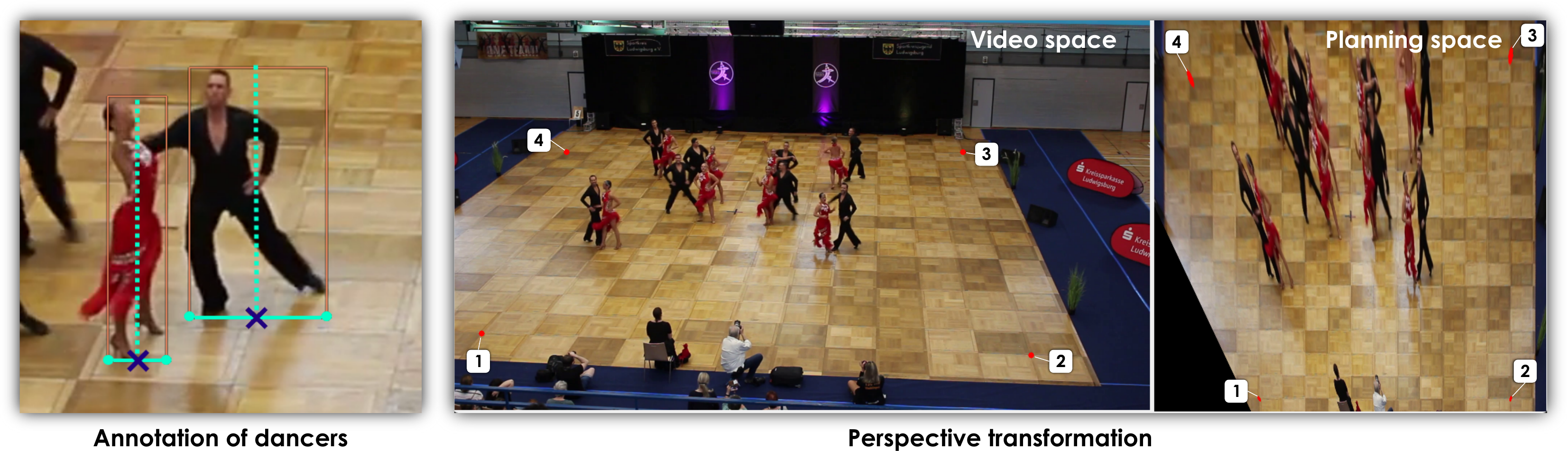}
    \caption{Bounding box annotation for individual dancers, the 2D position is determined by the center-bottom point of the bounding box. Coordinates are then transformed from video space to planning space for comparison with the planned formations.}
    \label{fig:processing_steps}
\end{figure*}

\subsection{Requirements}\label{subsec:requirements}

\renewcommand{\arraystretch}{1.1}
\begin{table*}
    \centering
    \caption{The requirements for our \emph{ChoreoVis} approach that were identified in the expert group interviews.}
    \begin{tabularx}{\textwidth}{| c | c | l | X |}
        \hline
        \multirow{7}{*}{\rotatebox[origin=c]{90}{\textbf{Plan}}} & R1 & Show Positions & A simple way to visualize the formations and see the dancers' positions in them. \\
        \cline{2-4}
        & R2 & Easy Editing & The possibility to easily create and edit formation definitions. \\
        \cline{2-4}
        & R3 & Show Orientations & Show the head and body orientations of the dancers in a formation. \\
        \cline{2-4}
        & R4 & Show Transitions & Visualize the transitions between formations. \\
        \cline{2-4}
        & R5 & Analyze Formations & The possibility to analyze the transitions and the choreography's utilization of the dance floor. \\
        \cline{2-4}
        & R6 & Show Poses & The possibility to view formation definitions in 3D to get a more realistic impression of the formation, including the dancers' poses. \\
        \hline
        \hline
        \multirow{4}{*}{\rotatebox[origin=c]{90}{\textbf{Assess}}} & R7 & Find Issues & An overview of the precision across the entire performance to quickly identify problematic parts. \\
        \cline{2-4}
        & R8 & Analyze Deviations & The possibility to analyze which dancers did not hit their positions in a formation, by how far they were off, and to communicate these findings to the respective dancers to aid their training sessions. \\
        \cline{2-4}
        & R9 & Validate Findings & Validating the findings made with the technique by linking them back to the video. \\
        \hline
    \end{tabularx}
    \label{tab:requirements}
\end{table*}

As mentioned before, we conducted multiple group interviews to gather requirements for our \emph{ChoreoVis} approach.
\autoref{tab:requirements} lists the requirements for an interactive formation planning prototype (R1 to R6) and for a prototype for visually assessing the realization of the formations in a dance performance (R7 to R9).

\section{ChoreoVis}\label{sec:choreoVis}

In the following, we present \emph{ChoreoVis}, our novel approach for interactive planning and assessing formations in dance choreographies developed to solve the tasks and fulfill the requirements described in \autoref{sec:technique}.
The approach is implemented as a web-based prototype using React and D3.js~\cite{bostock2011d3} with a responsive design to support mobile usage scenarios during training.
The planning component of the approach is actively used by our collaborators and continuously updated to incorporate their feedback.
We first discuss the data processing steps necessary for the approach before introducing the interactive planning prototype and, finally, the technique for visual assessment of choreographed dance performances.

\begin{figure*}[t]
    \centering
    \includegraphics[width=0.9\textwidth]{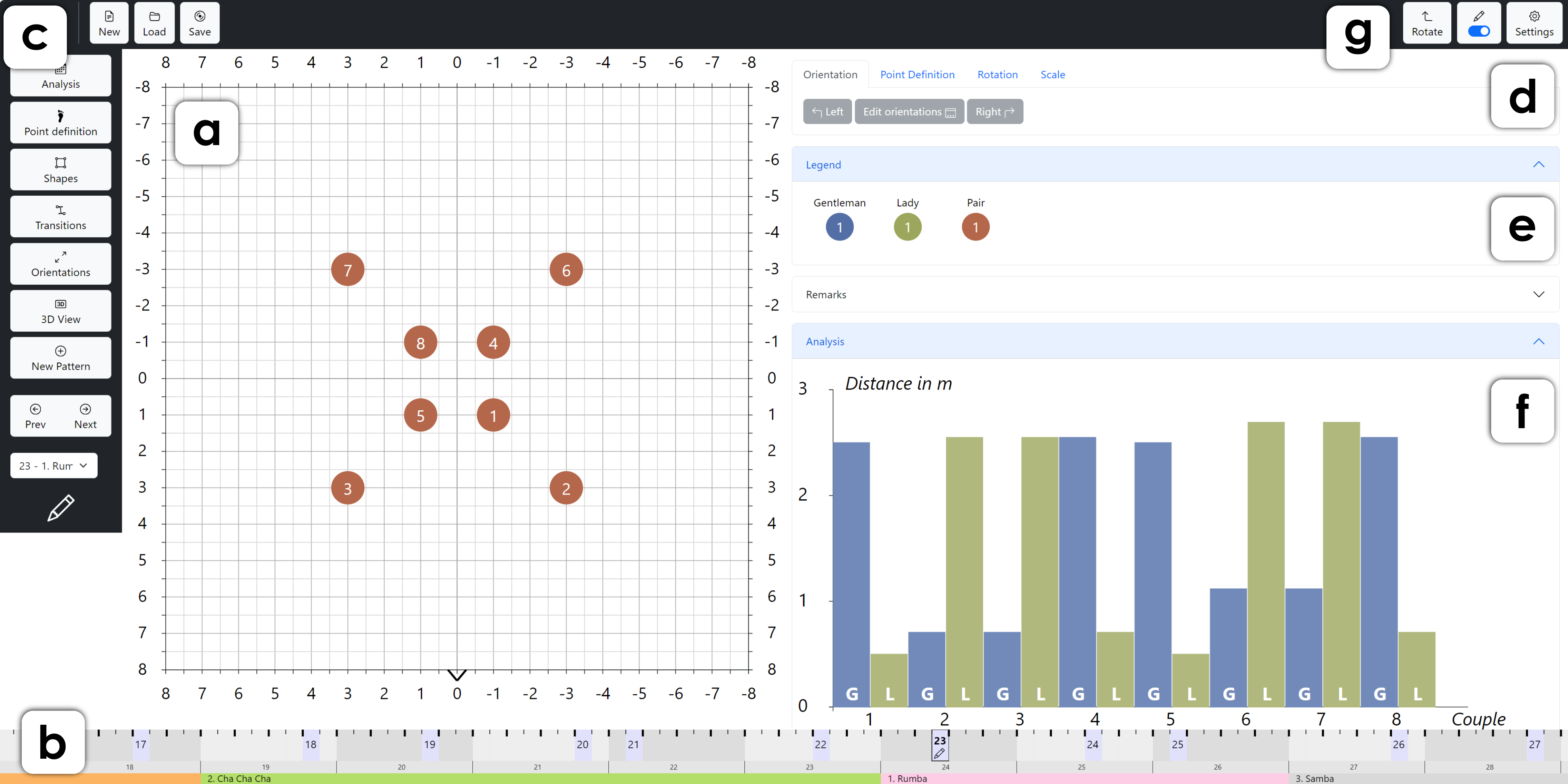}
    \caption{Interface of the choreography planning prototype: (a) the dance floor visualization shows the position of each dancer or couple in a formation, (b) a timeline shows the temporal arrangement of the formations, (c) a toolbar for changing views and navigating the formations, (d) a quick access menu for common edit tools, (e) the visualization legend, (f) a bar chart showing the distance moved between formations for each dancer, and (g) an app bar for loading and saving choreographies, switching between viewing and editing modes, and settings.}
    \label{fig:planning_interface}
\end{figure*}

\subsection{Data Processing}\label{subsec:dataprocessing}

The formations for planning are saved as JSON files.
Formation assessment additionally requires trajectory data of each dancer and a projection of coordinates from the performance to the planning coordinate system.

\paragraph*{Annotation of Dancers}
To obtain this data, we extracted the trajectories from the video of a dance performance.
An example of a danced choreography from our collaborators is discussed in \autoref{sec:case_study}.
The presented example was manually annotated with bounding boxes for each dancer.
We used a framework that was also applied to label moving objects in eye-tracking videos with keyframe annotation and interpolation in-between~\cite{kurzhals2014iseecube}. 
The resulting trajectories are saved in the form of an XML file that stores the position of each bounding box in each frame.

\paragraph*{Perspective Transformation}
The data at hand is located in the two coordinate systems: (1) the planning space and (2) the video space. Hence, a transformation from one space to the other has to be defined to make actual and planned trajectories comparable.
We used OpenCV~\cite{opencv_library} to project the positions of the bounding boxes in the video to positions in the virtual planning space. The transformation matrix was defined by correspondence points on the dancefloor and the grid in the planning space.

The resulting trajectories in planning space are defined by the bottom-center point of a bounding box per video frame (\autoref{fig:processing_steps}).
The formation definitions, trajectories, and transformed trajectories serve as input for our formation assessment technique.

\begin{figure*}[t]
    \centering
    \includegraphics[width=\textwidth]{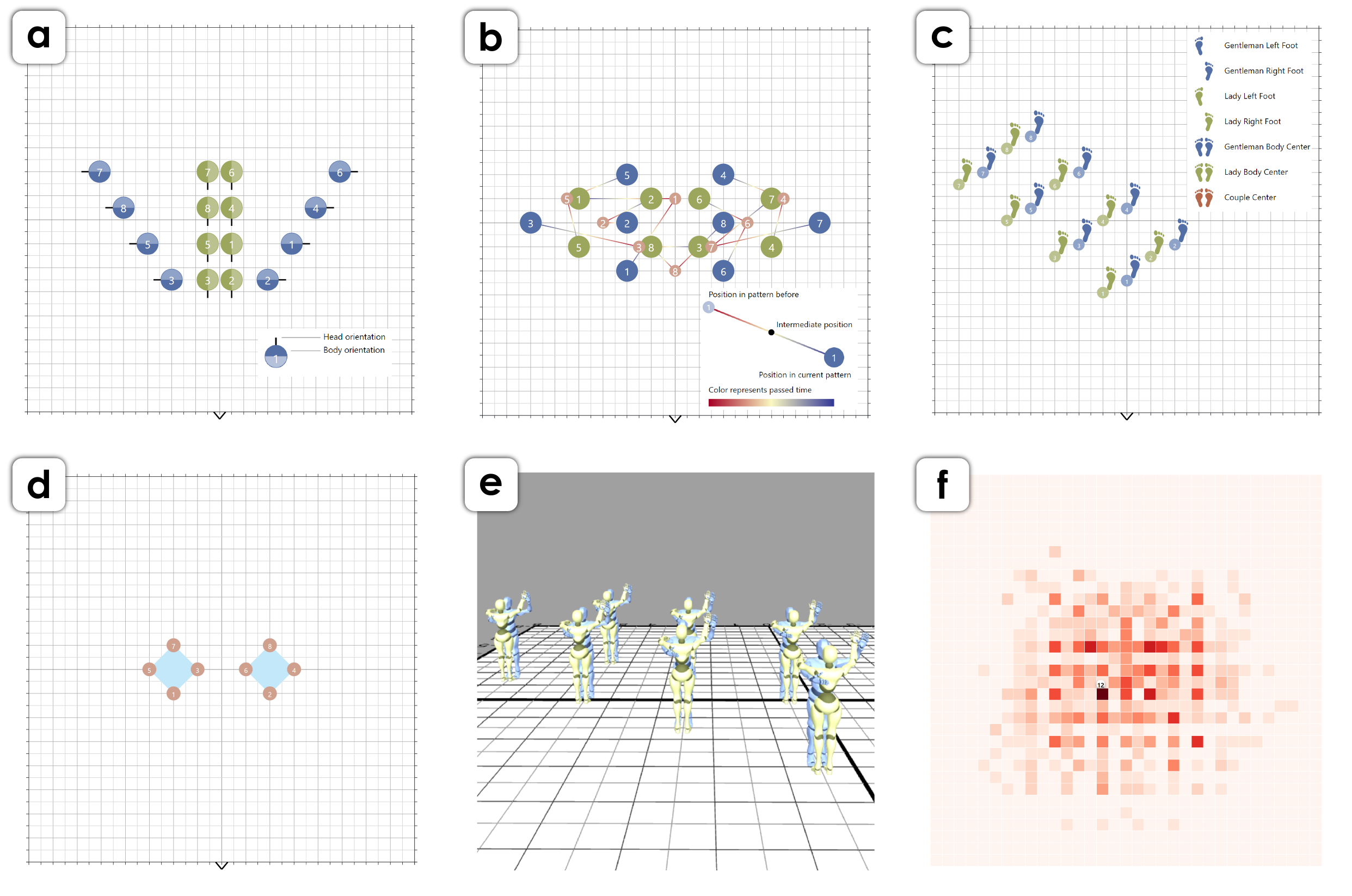}
    \caption{Included views: (a) the orientation view shows the position and orientation of each dancer in a formation, (b) the transition view shows the routes dancers take between formations and reveals potential problems, (c) the point definition view shows for which body part of which dancer in a couple the position is specified, (d) the shape view shows geometric abstractions of formations, (e) the 3D view shows dancers' poses, and (f) the heatmap shows the utilization of the dance floor.}
    \label{fig:planning_views}
\end{figure*}

\subsection{Visualization for Planning}\label{subsec:planning}

The first component of our \emph{ChoreoVis} approach covers the \emph{plan}, \emph{analyze}, and \emph{memorize} tasks of the workflow described in \autoref{fig:pattern-workflow}.
The intended user groups of the web-based prototype, shown in \autoref{fig:planning_interface}, are formation instructors and dancers.
It is divided into an editing mode for the instructors and a viewing mode for instructors and dancers designed primarily for mobile use during their practice sessions.
Editing is restricted to desktop browsers as it requires complex user interactions.
A visual representation of the dance floor (\autoref{fig:planning_interface}~(a)) 
shows the dancers' positions in the formations~(\emph{R1}).
The timeline (\autoref{fig:planning_interface}~(b)) shows the temporal arrangement of the formations.
A toolbar (\autoref{fig:planning_interface}~(c)) allows navigating between formations, adding new formations, and switching between views.
Next, we will go into detail about the different views that are part of the planning prototype.

\paragraph*{Dance Floor View}
The dance floor view (\autoref{fig:planning_interface}~(a)) consists of a Cartesian coordinate system representing the dance floor.
Dancers' and couples' positions are represented as dots; a small arrow indicates the front of the dance floor.
Our experts pointed out that the consistency, i.e., the spatial relationships between the dancers, is more important than the absolute positions on the dance floor. 
Our approach facilitates this style of training by highlighting all direct neighbors of a dancer or couple when hovering over them~(\emph{R1}). 
In the editing mode, users can move dancers or couples to a new position via drag and drop.
A brush can be used to select and edit multiple dancers and couples together. 
The quick access menu (\autoref{fig:planning_interface}~(d)) provides the ability to rotate selected lines with a slider~(\emph{R2}).

\paragraph*{Timeline}
The timeline (\autoref{fig:planning_interface}~(b)) shows the temporal arrangement of the formations.
Colored lines indicate the different dances that make up the choreography.
Each dance consists of multiple bars in the music that are reflected in the timeline by rectangles in alternating shades of gray.
This concept allows for precisely placing the formations in the choreography's music.
A small preview of a formation is shown when hovering over it to facilitate easy navigation.
The temporal position of a formation is adjustable. 

\paragraph*{Orientation View}
The dancers' head and body orientations are encoded through glyphs that are displayed in the orientation view (\autoref{fig:planning_views}~(a)).
The glyph consists of two semicircles with a lighter and a darker color, where the darker semicircle indicates the body orientation.
A black line indicates the orientation of the head, mimicking a nose~(\emph{R3}).
We arrived at this design after multiple iterations as it is intuitive to understand.
Users can change the orientation of one or multiple dancers using sliders~(\emph{R2}).

\paragraph*{Transition View}
The transition view shows the current formation, the previous formation, and the dancers' transitions between them in a single visualization (\autoref{fig:planning_views}~(b)).
The transitions are calculated as piece-wise linear functions as a best-effort estimation of the paths taken by the dancers.
They are represented by lines~(\emph{R4}) that are colored using the RdYlBu diverging color scheme to encode the temporal dimension of the transition.
This is particularly useful for identifying potential collisions of dancers already during the analysis process, recognizable by intersecting lines with similar color values~(\emph{R5}).
The viewing mode provides an alternative analysis method by animating the transitions in the dance floor view~(\emph{R4}). 
To specify transitions more precisely, users can add waypoints to them in the editing mode.
This is done by clicking on a bar in the timeline to specify at which point in the choreography the waypoint is defined.
Subsequently, the waypoint can be moved in the transition view via drag and drop to specify the transition~(\emph{R2}).
An important factor during the analysis of transitions is the distance that dancers have to move between formations.
For this reason, a bar chart (\autoref{fig:planning_interface}~(f)) accompanies the transition view and visualizes the length of each dancer's path in meters~(\emph{R5}).

\paragraph*{Point Definition View}
The point definition view (\autoref{fig:planning_views}~(c)) gives a more precise definition of which dancer of a dance couple and which part of a dancer's body is standing on the defined point in the formation~(\emph{R1}).
The options for this definition are the center between a dance couple and the body center, right foot, or left foot of a dancer.
We use foot-shaped glyphs as an intuitive encoding for this information in the view.
A left or right foot pictogram represents the respective information, while two feet represent the center.
Color is used to encode the dancer in a couple that is standing at the defined position.
In the edit mode, clicking on a dancer or dance couple allows selecting the body part from a list of options~(\emph{R2}).

\paragraph*{Shape View}
The goal of the shape view (\autoref{fig:planning_views}~(d)) is to improve the memorability of formations by providing an abstraction of them.
This directly supports the \emph{memorize} task.
These abstractions are geometric shapes, which can be basic building blocks of formations, such as diamonds or rectangles.
They are defined by the convex hull of a user-created selection of dancers.
We decided against the automatic recognition of shapes as dance formations contain many different shapes that do not all boost memorability.

\paragraph*{3D View}
A three-dimensional view (\autoref{fig:planning_views}~(e)) is intended to show the poses of the dancers in a formation~(\emph{R6}).
We decided to use a 3D visualization to avoid having users learn complex glyph designs.
Thus, each dancer is represented by a three-dimensional avatar on the dance floor at the position and with the orientation from the other views.
In the editing mode, a pose editor is shown on the right side of the interface.
Here, users can select joints from a drop-down menu and change their rotation using a transform control directly on the avatar to create complex poses.
To avoid repetitive work, the pose can be applied to multiple dancers at once~(\emph{R2}).

\paragraph*{Analysis View}
The analysis view (\autoref{fig:planning_views}~(f)) aims to give insights into the choreography with a heatmap that provides an overview of the utilization of the dance floor throughout the choreography, where the color of a position indicates how often a dancer is standing on it in a formation~(\emph{R5}).
The positions during transitions are not considered in the heatmap as they are not an essential quality metric for the choreography. 
Next to it is a bar chart (\autoref{fig:planning_interface}~(f)) that plots the dancers' accumulated movement distances over the entire choreography.

\begin{figure*}[t]
    \centering
    \includegraphics[width=0.95\textwidth]{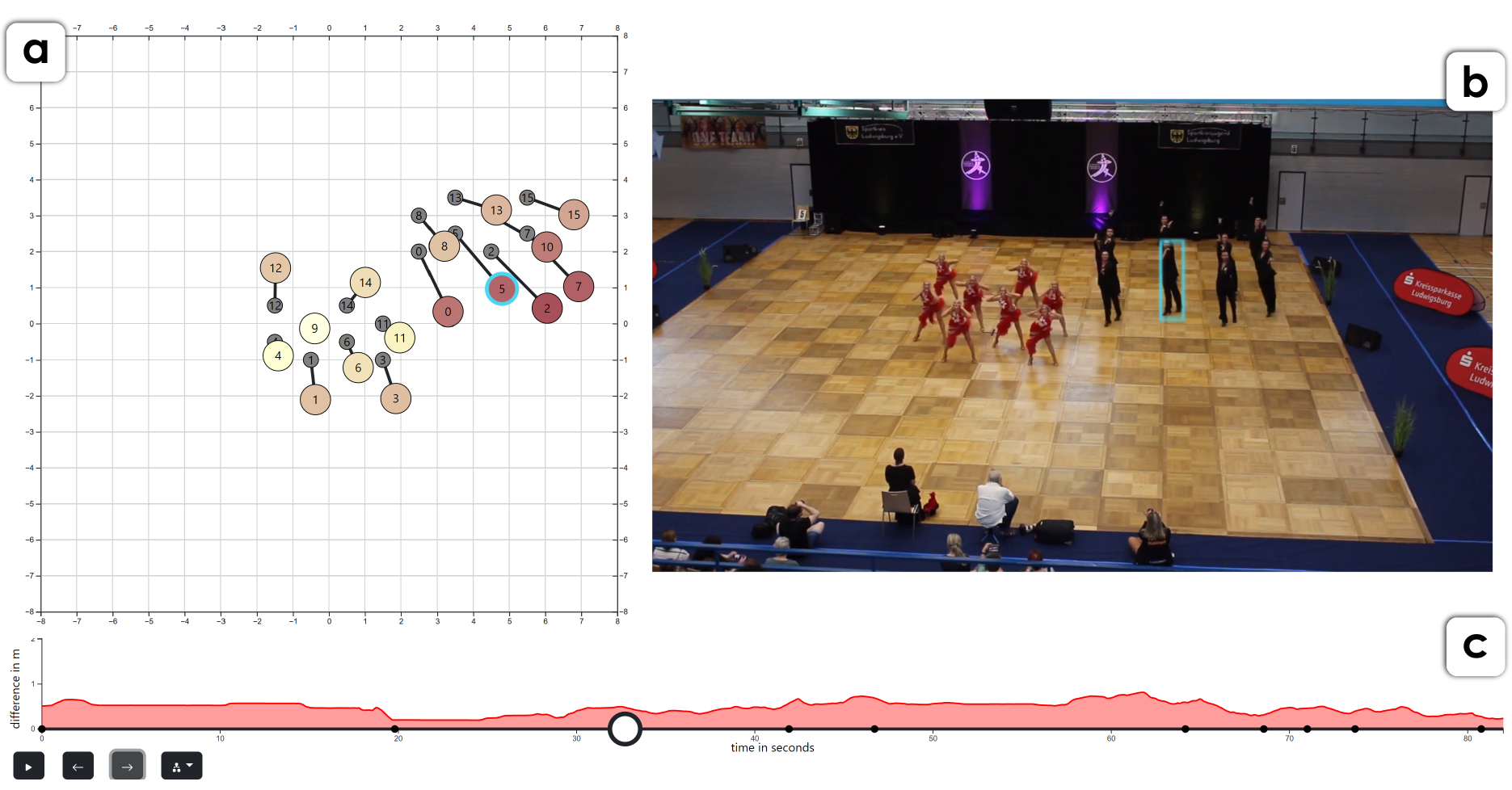}
    \caption{Interface of the choreography assessment prototype: (a) the dance floor view shows the difference between defined and danced positions at a specific time, (b) a video of the performance with the bounding boxes of selected dancers, and (c) a timeline that visualizes the difference between defined and danced positions over the entire choreography.}
    \label{fig:assessing_interface}
\end{figure*}

\subsection{Visual Assessment of Dance Performances}\label{subsec:assessing}

In current practice, instructors assess performances typically by manual video analysis without additional support through visualization techniques. To improve this procedure, we link the video of the performance with the planning view (\autoref{fig:assessing_interface}). We apply a derived perspective transformation to the bounding box data from the video (\autoref{subsec:dataprocessing}), which allows displaying the video positions in a similar way as the planned positions.

\paragraph*{Dance Floor View}
For a selected time step, the abstract representation shows the planned position as small circles and the actual position as big circles connected with a line (\autoref{fig:assessing_interface}~(a)) to help assess the deviation from the defined formation~(\emph{R8}).
We use the YlOrRd color scale to encode the deviation additionally by color. Correct positions are indicated in yellow, while strong deviations are displayed in red.
Individual dancers can be selected and are also highlighted in the video.

\paragraph*{Video}
The video (\autoref{fig:assessing_interface}~(b)) is an integral part of the analysis~(\emph{R9}) and will be assessed in similar ways by the instructor as defined by established practices. However, the visualization supports identifying individual problems and keeping track of dancers.
Selected dancers are highlighted, and their performance over time can be investigated with the difference timeline.

\paragraph*{Difference Timeline}
On the timeline, we encode the mean square error of deviations from the original position over time (\autoref{fig:assessing_interface}~(c)). The timeline shows calculations based on selected dancers, allowing to assess the whole team, couples, and individuals. Additionally, the individual dance formations are marked on the timeline to ease the identification of difficult parts in the choreography~(\emph{R7}).

\section{Case Study}\label{sec:case_study}

To demonstrate the applicability of our \emph{ChoreoVis} approach, we describe a case study that covers the workflow's \emph{plan}, \emph{analyze}, and \emph{assess} stages based on a real formation choreography.
Our prototypes and the described data can be found in the supplementary material.

\subsection{Data}

The data used in the case study is from a Latin formation choreography adapted and performed at competitions by the formation team we are collaborating with.
The choreography lasts five minutes and 50 seconds and comprises 57 formations.
Our collaborators provided us with the formation definitions in the form of a PDF file and a video recording of a performance at a formation competition.
The video was recorded from an elevated position in front of the dance floor and is slightly off-center.
The front left corner of the dance floor is not visible in the video, which is negligible as that area is not used in the performance.

\subsection{Choreography Planning and Analysis}

Based on the definitions in the PDF file, we recreated all 57 formations using our interactive planning approach.
Thanks to the easy positioning of the dancers via drag and drop and the possibility of duplicating formations before applying minor changes to them, this process was fast and straightforward.
We made our prototype and the recreated choreography available to our collaborators for them to use during training sessions and to analyze and memorize the formation definitions.
The team used the prototype frequently for eight months to look up positions during practice and make minor adjustments to some of the formations.
It was received enthusiastically by dancers and instructors, who also gave frequent feedback, which we used to improve the prototype, like adding the possibility to use the arrow keys for navigation.

Subsequently, we discuss the applicability of the developed views to some of the cases in the choreography.
\autoref{fig:planning_views}~(a) shows how the orientation view gives a quick overview of head and body orientations in a formation.
In this example, the ladies form two lines in the middle of the dance floor.
The glyphs indicate that they look towards the front while their bodies are oriented to the right.
Meanwhile, the gentlemen form two diagonals, their bodies oriented forward in the pose while their heads look outward.
The formation transition view shows that the path of Lady 6 and Gentleman 7 are crossing during the transition in \autoref{fig:planning_views}~(b).
The lines having almost the same color at their intersection indicate a potential issue in the transition that should be addressed when training the respective passage.
Indeed, this problem came up in practice and was resolved by Lady 6 waiting for Gentleman 7 to pass first before moving to the front since the distance she had to cover was much shorter.
The distance bar chart reveals that she has to move only 2.2 meters while the other dancer has to move 4.6 meters with the same steps in the same amount of time, making such a fix reasonable.
To form precise lines in a formation, dancers must have the same understanding of a point definition.
In wide-legged poses, it can make a drastic difference which foot dancers put on the defined point.
As seen in \autoref{fig:planning_views}~(c), the point definition view clarifies this matter.
In this example, all dancers must put their right foot on the point to form straight diagonals.
Regarding the dance floor utilization, the heatmap in \autoref{fig:planning_views}~(f) indicates that the choreography utilizes the dance floor uniformly without any striking differences.

\subsection{Performance Assessment}

To extract the dancer trajectories from the video, we annotated the first 82 seconds of the choreography with bounding boxes around each dancer using the annotation tool introduced by Kurzhals et al.~\cite{kurzhals2014iseecube}.
As the manual annotation is a time-intensive task, we deemed it sufficient to annotate only part of the video (covering ten formations) as this provides enough data for the purpose of this case study.
We then transformed the resulting trajectory data from the video space to the planning space with OpenCV, as illustrated in \autoref{fig:processing_steps}, using four points in the corners of the dance floor as reference points.
Additionally, we performed a linear interpolation of the discrete formation definitions created with our planning prototype to obtain intermediate baseline data for each video frame.
To compute the interpolation, we added a time stamp to each formation definition corresponding to its appearance in the video.
When the interpolated formation definitions, the trajectories, and the transformed trajectories are opened in our assessment prototype, users are presented with the dance floor view, video, and timeline as shown in \autoref{fig:assessing_interface}.

In the following, we discuss interesting cases in our dataset.
The timeline shows that the average mean squared error varies between 0.2 and 0.8 meters, with the lowest error at the second formation and the highest shortly before the sixth formation.
\autoref{fig:assessing_interface} shows the third choreography formation with a mean error of around 0.5 meters.
In the dance floor view, the gentlemen in the front are colored dark red to indicate a significant deviation from the definition.
Upon taking a closer look, it is clearly recognizable that they are standing around one meter further to the right and around one-and-a-half meters further to the front than defined in the formation.
This finding is indicated by the connecting lines between the actual positions and the definitions and can be confirmed in the video view.
Furthermore, the visualization indicates that Dancer 1 and Dancer 3 are one meter ahead of their defined positions.
Such mistakes are hard to spot without our approach since the danced formation looks mostly consistent despite the deviations.
However, with the timeline and the color encoding, they are easy to identify with our approach.

Another interesting case is the tenth formation shown in \autoref{fig:teaser}.
Here, the couples have mostly reached their positions in the formation except for Dancers 8 and 9 and Dancers 12 and 13 on the left side, which are still transitioning to their new positions from the previous formation.
Thus, our prototype is also suited for finding transitions that are not executed uniformly and disturb the coherent impression of the choreography.
Finding such nonuniform transitions is a challenging task with only the video recording.

Finally, the first formation demonstrates why the video view is essential for validating the findings made in the other views.
The dance floor view shows the dancers in the back deviating significantly from their positions.
However, upon inspecting the video, it becomes clear that the dancers hit their positions, and the fault is with the bounding boxes.
Since the bottom half of the dancers in the back is occluded from sight by the dancers standing before them, the annotator had to estimate the sizes of the bounding boxes.
Since they are too large, the trajectories are skewed slightly, explaining the significant deviation shown in the other views.

\section{Expert Study}\label{sec:evaluation}

We conducted a think-aloud study~\cite{boren2000thinking} to collect feedback about an earlier version of our interactive planning approach that was still missing the point definition view, the timeline's separation into dances, and some usability features.
The participants of the study were four formation instructors who had between 1.5 and 21 years of instructing experience across all levels of competitive formation dancing in Germany.
Three of the participants were not involved in the design process of the prototype, and all had previous experience with tools for planning choreographies.
In sessions taking between 60 and 90 minutes, the participants had to solve 19 tasks pertaining to the \emph{plan} and \emph{analyze} steps of our workflow while verbalizing their thought processes.
They involved understanding the visualizations, creating and editing formations, transitions, and orientations, and analyzing a predefined choreography.
The data used in the study were unknown to the participants. They had a 15-minute training phase to familiarize themselves with the prototype.
Afterward, they rated different aspects of the prototype on a Likert scale from -2 to 2.
Below, we present the results of our analysis of the participants' interactions with the prototype, the study transcripts, and the questionnaire.
The task descriptions and transcripts can be found in the supplementary material.

The participants rated the visualizations as easy to understand, with a median of 2.
They were able to identify positions, orientations, and poses without major problems.
Furthermore, they thought creating new formations and arranging them on the timeline was simple and fast, with a median of 1.5.
The provided template formations and the feature to create custom templates were praised as unique to our approach.
The participants regarded the separation into multiple views and the straightforward user interface especially positively.
Two participants highlighted the easy detection of potential collisions in the transition view, which is not possible with existing tools.
Additionally, the participants praised the analysis views, like the distance bar chart and the heatmap, as unique features missing in other tools.
Similarly, three participants commented that the 3D view offers novel and exciting features that add value besides the 2D views.
An aspect that was criticized in the study was the fixed division of a music bar in the timeline into eight beats, as this is not universally applicable to all dances.
One of the participants suggested replacing bars in the timeline with dances.
Likewise, it was doubted if the shape view is helpful since the shape of a formation is usually recognizable without a visual indication.
Overall, the participants were enthusiastic about the developed approach and could solve the given tasks without major problems.
We incorporated much of the feedback into the version of the approach that is presented in \autoref{subsec:planning}.

\section{Discussion}\label{sec:discussion}

Based on the feedback from domain experts, we could identify some limitations of the approach. Overall, the approach was received well, but some aspects will require further research.

\paragraph*{Dance Couple Tracking}
Our current implementation showed that with appropriate bounding boxes, an assessment of the quality of the choreography is possible, and instructors can identify problematic events effectively. However, the consistent tracking of dance couples poses some practical challenges to state-of-the-art computer vision techniques, as uniform dresses, occlusions, or insufficient video quality might impair a fully automatic extraction of the data. We are confident that current approaches for tracking (e.g., Doersch et al.~\cite{doersch2023tapir}) in combination with specifically trained dance couple detectors could provide sufficient results, eventually combined with a semi-automatic labeling process which could also be integrated into the existing visualization framework.

\paragraph*{Abstract vs. Detailed Planning and Assessment}
With the proposed planning approach, we make deliberate choices regarding the level of detail that can be visually represented for a choreography.
On the one hand, it is important to come up with initial formations quickly, which requires fast and abstract interactive visual means to represent them.
On the other hand, many details, such as poses and moves of individual dancers and couples that influence the quality of a choreography, cannot be modeled at a fine-granular level.
Considering the current technical setup our collaborating practitioners have access to, we believe that our solution is close to a sweet spot regarding necessary abstraction and enough detail for utilizing the planned choreographies effectively and efficiently for training.
The feedback from the experts supports this assessment.
Given enough resources, it would certainly be possible to add more details quickly by tracking dancing couples’ movements and using these as prototypes, which could be incorporated as part of the planning.
This would also make it possible to assess the training progress at a much more fine-grained level, given such advanced tracking equipment was available.

\paragraph*{Other Application Domains}
We argue that \emph{ChoreoVis} can be applied to any choreographed group activity that includes spatiotemporal arrangements.
Examples include other group dances like hip-hop or cheerleading, marching band formations, and theater and musical performances.
However, additional research is required to confirm this hypothesis.

\section{Conclusion}\label{sec:conclusion}

We presented \textit{ChoreoVis}, a visualization approach for planning and assessing dance choreographies, using the example of Latin formation dancing.
We also discussed the tasks and workflows of planning, practicing, and assessing formations and the requirements that we derived from them for our approach.
With an annotated recording of the dance performance, instructors can assess the performance of individual couples and provide targeted instructions for improvement. In contrast to the regular assessment by watching the video, the visualization provides hints where large deviations from the defined choreography occur and supports more efficient training.
In the future, we plan to address additional scenarios based on choreographed performances. One remaining issue to achieve this is the individual differences in required computer vision algorithms to acquire the data for the visualization. We based our technique on manually annotated data for optimal tracking and semantic coherence of dance couples. Automatic detection and recognition of dance couples for tracking is desirable but poses a separate research problem that was not the focus of this work.
Professional dancers and instructors acknowledged the use of visualization and visual analytics for choreography training very positively.
With future extensions like immersive analytics techniques that also support the practice step after feedback from the assessment, training methods could be extended to teach new choreographies more efficiently.
For now, our collaboration partners successfully adopted the planning component of our approach in practice to plan and learn their new choreography.

\section*{Acknowledgments}
We thank the participants of the expert study and TSC fun\&dance Waiblingen e.V. for the excellent collaboration.
[Partially] Funded by Deutsche Forschungsgemeinschaft (DFG, German Research Foundation) under Germany’s Excellence Strategy - EXC 2075 – 390740016 and EXC 2120/1 – 390831618.
We acknowledge the support by the Stuttgart Center for Simulation Science (SimTech).

\bibliographystyle{eg-alpha-doi} 
\bibliography{bibliography}       

\newcommand{\etalchar}[1]{$^{#1}$}
\begin{thebibliography}{\uppercase{MRKEA22}}

\bibitem[AGH{\etalchar{*}}23]{afzal2023visualization}
\textsc{Afzal S., Ghani S., Hittawe M.~M., Rashid S.~F., Knio O.~M., Hadwiger M., Hoteit I.}:
\newblock Visualization and visual analytics approaches for image and video datasets: A survey.
\newblock \emph{ACM Transactions on Interactive Intelligent Systems 13}, 1 (2023), 1--41.
\newblock \href {https://doi.org/10.1145/3576935} {\path{doi:10.1145/3576935}}.

\bibitem[AMST11]{aigner2011visualization}
\textsc{Aigner W., Miksch S., Schumann H., Tominski C.}:
\newblock \emph{Visualization of time-oriented data}, vol.~4.
\newblock Springer, 2011.
\newblock \href {https://doi.org/10.1007/978-1-4471-7527-8} {\path{doi:10.1007/978-1-4471-7527-8}}.

\bibitem[BCD{\etalchar{*}}12]{borgo2012state}
\textsc{Borgo R., Chen M., Daubney B., Grundy E., Heidemann G., H{\"o}ferlin B., H{\"o}ferlin M., Leitte H., Weiskopf D., Xie X.}:
\newblock State of the art report on video-based graphics and video visualization.
\newblock \emph{Computer Graphics Forum 31}, 8 (2012), 2450--2477.
\newblock \href {https://doi.org/10.1111/j.1467-8659.2012.03158.x} {\path{doi:10.1111/j.1467-8659.2012.03158.x}}.

\bibitem[BDS{\etalchar{*}}23]{beck2023visual}
\textsc{Beck S., Doerr N., Schmierer F., Sedlmair M., Koch S.}:
\newblock Visual planning and analysis of latin formation dance patterns.
\newblock In \emph{EuroVis 2023 - Posters} (2023), Gillmann C., Krone M., Lenti S., (Eds.), The Eurographics Association.
\newblock \href {https://doi.org/10.2312/evp.20231076} {\path{doi:10.2312/evp.20231076}}.

\bibitem[BLB{\etalchar{*}}16]{brehmer2016timelines}
\textsc{Brehmer M., Lee B., Bach B., Riche N.~H., Munzner T.}:
\newblock Timelines revisited: A design space and considerations for expressive storytelling.
\newblock \emph{IEEE Transactions on Visualization and Computer Graphics 23}, 9 (2016), 2151--2164.
\newblock \href {https://doi.org/10.1109/TVCG.2016.2614803} {\path{doi:10.1109/TVCG.2016.2614803}}.

\bibitem[BOH11]{bostock2011d3}
\textsc{Bostock M., Ogievetsky V., Heer J.}:
\newblock D³ data-driven documents.
\newblock \emph{IEEE Transactions on Visualization and Computer Graphics 17}, 12 (2011), 2301--2309.
\newblock \href {https://doi.org/10.1109/TVCG.2011.185} {\path{doi:10.1109/TVCG.2011.185}}.

\bibitem[BR00]{boren2000thinking}
\textsc{Boren T., Ramey J.}:
\newblock Thinking aloud: {R}econciling theory and practice.
\newblock \emph{IEEE Transactions on Professional Communication 43}, 3 (2000), 261--278.
\newblock \href {https://doi.org/10.1109/47.867942} {\path{doi:10.1109/47.867942}}.

\bibitem[Bra00]{opencv_library}
\textsc{Bradski G.}:
\newblock {The OpenCV Library}.
\newblock \emph{Dr. Dobb's Journal of Software Tools} (2000).

\bibitem[CLX{\etalchar{*}}16]{chen2016}
\textsc{Chen W., Lao T., Xia J., Huang X., Zhu B., Hu W., Guan H.}:
\newblock Gameflow: Narrative visualization of nba basketball games.
\newblock \emph{IEEE Transactions on Multimedia 18}, 11 (2016), 2247--2256.
\newblock \href {https://doi.org/10.1109/TMM.2016.2614221} {\path{doi:10.1109/TMM.2016.2614221}}.

\bibitem[CM21]{coenen2021public}
\textsc{Coenen J., Moere A.~V.}:
\newblock Public data visualization: {A}nalyzing local running statistics on situated displays.
\newblock \emph{Computer Graphics Forum 40}, 3 (2021), 159--171.
\newblock \href {https://doi.org/10.1111/cgf.14297} {\path{doi:10.1111/cgf.14297}}.

\bibitem[DKVS14]{dietrich2014baseball4d}
\textsc{Dietrich C., Koop D., Vo H.~T., Silva C.~T.}:
\newblock Baseball{4D}: {A} tool for baseball game reconstruction \& visualization.
\newblock In \emph{IEEE Conference on Visual Analytics Science and Technology (VAST)} (2014), pp.~23--32.
\newblock \href {https://doi.org/10.1109/VAST.2014.7042478} {\path{doi:10.1109/VAST.2014.7042478}}.

\bibitem[DTW{\etalchar{*}}13]{duffy2013glyph}
\textsc{Duffy B., Thiyagalingam J., Walton S., Smith D.~J., Trefethen A., Kirkman-Brown J.~C., Gaffney E.~A., Chen M.}:
\newblock Glyph-based video visualization for semen analysis.
\newblock \emph{IEEE Transactions on Visualization and Computer Graphics 21}, 8 (2013), 980--993.
\newblock \href {https://doi.org/10.1109/TVCG.2013.265} {\path{doi:10.1109/TVCG.2013.265}}.

\bibitem[DY21]{du2021survey}
\textsc{Du M., Yuan X.}:
\newblock A survey of competitive sports data visualization and visual analysis.
\newblock \emph{Journal of Visualization 24} (2021), 47--67.
\newblock \href {https://doi.org/10.1007/s12650-020-00687-2} {\path{doi:10.1007/s12650-020-00687-2}}.

\bibitem[DYV{\etalchar{*}}23]{doersch2023tapir}
\textsc{Doersch C., Yang Y., Vecerik M., Gokay D., Gupta A., Aytar Y., Carreira J., Zisserman A.}:
\newblock Tapir: Tracking any point with per-frame initialization and temporal refinement.
\newblock In \emph{2023 IEEE/CVF International Conference on Computer Vision (ICCV)} (2023), pp.~10027--10038.
\newblock \href {https://doi.org/10.1109/ICCV51070.2023.00923} {\path{doi:10.1109/ICCV51070.2023.00923}}.

\bibitem[{Ger}98]{DTV1998}
\textsc{{German Dancing Federation (DTV)}}:
\newblock Wertungsrichtlinien im {DTV} für formationswettbewerbe standard und latein.
\newblock \url{https://www.tanzsport.de/files/tanzsport/downloads/sportwelt/formationen/wrichtl-f.pdf}, 1998.

\bibitem[GZX{\etalchar{*}}22]{guo2022dancevis}
\textsc{Guo H., Zou S., Xu Y., Yang H., Wang J., Zhang H., Chen W.}:
\newblock Dancevis: Toward better understanding of online cheer and dance training.
\newblock \emph{Journal of Visualization 25}, 1 (2022), 159--174.
\newblock \href {https://doi.org/10.1007/s12650-021-00783-x} {\path{doi:10.1007/s12650-021-00783-x}}.

\bibitem[HHHW13]{hoeferlin2013}
\textsc{Hoeferlin M., Hoeferlin B., Heidemann G., Weiskopf D.}:
\newblock Interactive schematic summaries for faceted exploration of surveillance video.
\newblock \emph{IEEE Transactions on Multimedia 15}, 4 (2013), 908--920.
\newblock \href {https://doi.org/10.1109/TMM.2013.2238521} {\path{doi:10.1109/TMM.2013.2238521}}.

\bibitem[HHHW15]{hoferlin2015scalable}
\textsc{H{\"o}ferlin B., H{\"o}ferlin M., Heidemann G., Weiskopf D.}:
\newblock Scalable video visual analytics.
\newblock \emph{Information Visualization 14}, 1 (2015), 10--26.
\newblock \href {https://doi.org/10.1177/1473871613488571} {\path{doi:10.1177/1473871613488571}}.

\bibitem[{HOL}23]{formi}
\textsc{{HOLYFEET, LLC.}}:
\newblock {FORMI}: The collaborative choreography design tool.
\newblock \url{https://www.formistudio.app/}, 2023.
\newblock Last accessed on 2023-11-30.

\bibitem[HS22]{heyen2022augmented}
\textsc{Heyen F., Sedlmair M.}:
\newblock Augmented reality visualization for musical instrument learning.
\newblock In \emph{International Society for Music Information Retrieval Conference (ISMIR) Late-Breaking Demo} (2022).

\bibitem[KHW14]{kurzhals2014iseecube}
\textsc{Kurzhals K., Heimerl F., Weiskopf D.}:
\newblock {ISeeCube}: Visual analysis of gaze data for video.
\newblock In \emph{Proceedings of the Symposium on Eye Tracking Research and Applications} (2014), pp.~43--50.
\newblock \href {https://doi.org/10.1145/2578153.2578158} {\path{doi:10.1145/2578153.2578158}}.

\bibitem[KKM{\etalchar{*}}20]{khulusi2020survey}
\textsc{Khulusi R., Kusnick J., Meinecke C., Gillmann C., Focht J., J{\"a}nicke S.}:
\newblock A survey on visualizations for musical data.
\newblock In \emph{Computer Graphics Forum} (2020), vol.~39, pp.~82--110.
\newblock \href {https://doi.org/10.1111/cgf.13905} {\path{doi:10.1111/cgf.13905}}.

\bibitem[LCB{\etalchar{*}}23]{lin2023ball}
\textsc{Lin T., Chen Z., Beyer J., Wu Y., Pfister H., Yang Y.}:
\newblock The ball is in our court: Conducting visualization research with sports experts.
\newblock \emph{IEEE Computer Graphics and Applications 43}, 1 (2023), 84--90.
\newblock \href {https://doi.org/10.1109/MCG.2022.3222042} {\path{doi:10.1109/MCG.2022.3222042}}.

\bibitem[LTB16]{Losada2016}
\textsc{Losada A.~G., Therón R., Benito A.}:
\newblock {BKViz}: A basketball visual analysis tool.
\newblock \emph{IEEE Computer Graphics and Applications 36}, 6 (2016), 58--68.
\newblock \href {https://doi.org/10.1109/MCG.2016.124} {\path{doi:10.1109/MCG.2016.124}}.

\bibitem[MI13]{meghdadi2013interactive}
\textsc{Meghdadi A.~H., Irani P.}:
\newblock Interactive exploration of surveillance video through action shot summarization and trajectory visualization.
\newblock \emph{IEEE Transactions on Visualization and Computer Graphics 19}, 12 (2013), 2119--2128.
\newblock \href {https://doi.org/10.1109/TVCG.2013.168} {\path{doi:10.1109/TVCG.2013.168}}.

\bibitem[MRKEA22]{miller2022corpusvis}
\textsc{Miller M., Rauscher J., Keim D.~A., El-Assady M.}:
\newblock {CorpusVis}: Visual analysis of digital sheet music collections.
\newblock In \emph{Computer Graphics Forum} (2022), vol.~41, pp.~283--294.
\newblock \href {https://doi.org/10.1111/cgf.14540} {\path{doi:10.1111/cgf.14540}}.

\bibitem[MTM23]{menzel2023beat}
\textsc{Menzel M., Tauscher J.-P., Magnor M.}:
\newblock {On the Beat: Analysing and Evaluating Synchronicity in Dance Performances}.
\newblock In \emph{Vision, Modeling, and Visualization} (2023), Guthe M., Grosch T., (Eds.), The Eurographics Association.
\newblock \href {https://doi.org/10.2312/vmv.20231230} {\path{doi:10.2312/vmv.20231230}}.

\bibitem[Muh09]{muhammad2009}
\textsc{Muhammad M.~N.}:
\newblock \emph{Visualizing Dance Formations: The Choreographer’s Tool}.
\newblock Bachelor's thesis, Bryn Mawr College, 2009.

\bibitem[NWVG19]{napolean2019running}
\textsc{Napolean Y., Wibowo P.~T., Van~Gemert J.~C.}:
\newblock Running event visualization using videos from multiple cameras.
\newblock In \emph{Proceedings of the International Workshop on Multimedia Content Analysis in Sports} (2019), pp.~82--90.
\newblock \href {https://doi.org/10.1145/3347318.3355528} {\path{doi:10.1145/3347318.3355528}}.

\bibitem[OBCT24]{offenwanger2023timesplines}
\textsc{Offenwanger A., Brehmer M., Chevalier F., Tsandilas T.}:
\newblock {TimeSplines}: Sketch-based authoring of flexible and idiosyncratic timelines.
\newblock \emph{IEEE Transactions on Visualization and Computer Graphics 30}, 1 (2024), 34--44.
\newblock \href {https://doi.org/10.1109/TVCG.2023.3326520} {\path{doi:10.1109/TVCG.2023.3326520}}.

\bibitem[PJHY19]{polk2019}
\textsc{Polk T., J{\"a}ckle D., H{\"a}u{\ss}ler J., Yang J.}:
\newblock {CourtTime}: Generating actionable insights into tennis matches using visual analytics.
\newblock \emph{IEEE Transactions on Visualization and Computer Graphics 26}, 1 (2019), 397--406.
\newblock \href {https://doi.org/10.1109/TVCG.2019.2934243} {\path{doi:10.1109/TVCG.2019.2934243}}.

\bibitem[PM06]{page2006}
\textsc{Page M., Moere A.}:
\newblock Towards classifying visualization in team sports.
\newblock In \emph{International Conference on Computer Graphics, Imaging and Visualisation (CGIV'06)} (2006), pp.~24--29.
\newblock \href {https://doi.org/10.1109/CGIV.2006.85} {\path{doi:10.1109/CGIV.2006.85}}.

\bibitem[POJC01]{pingali2001}
\textsc{Pingali G., Opalach A., Jean Y., Carlbom I.}:
\newblock Visualization of sports using motion trajectories: Providing insights into performance, style, and strategy.
\newblock In \emph{Proceedings of IEEE Visualization (VIS)} (2001), pp.~75--82.
\newblock \href {https://doi.org/10.1109/VISUAL.2001.964496} {\path{doi:10.1109/VISUAL.2001.964496}}.

\bibitem[PVS{\etalchar{*}}18]{perin2018}
\textsc{Perin C., Vuillemot R., Stolper C.~D., Stasko J.~T., Wood J., Carpendale S.}:
\newblock State of the art of sports data visualization.
\newblock In \emph{Computer Graphics Forum} (2018), vol.~37, pp.~663--686.
\newblock \href {https://doi.org/10.1111/cgf.13447} {\path{doi:10.1111/cgf.13447}}.

\bibitem[PYHZ14]{polk2014tennivis}
\textsc{Polk T., Yang J., Hu Y., Zhao Y.}:
\newblock Tennivis: Visualization for tennis match analysis.
\newblock \emph{IEEE Transactions on Visualization and Computer Graphics 20}, 12 (2014), 2339--2348.
\newblock \href {https://doi.org/10.1109/TVCG.2014.2346445} {\path{doi:10.1109/TVCG.2014.2346445}}.

\bibitem[RHWS22]{rau2022visualization}
\textsc{Rau S., Heyen F., Wagner S., Sedlmair M.}:
\newblock Visualization for ai-assisted composing.
\newblock In \emph{International Society for Music Information Retrieval Conference (ISMIR)} (2022).

\bibitem[RLB19]{ren2018charticulator}
\textsc{Ren D., Lee B., Brehmer M.}:
\newblock {Charticulator}: Interactive construction of bespoke chart layouts.
\newblock \emph{IEEE Transactions on Visualization and Computer Graphics 25}, 1 (2019), 789--799.
\newblock \href {https://doi.org/10.1109/TVCG.2018.2865158} {\path{doi:10.1109/TVCG.2018.2865158}}.

\bibitem[SI11]{saito2011musicube}
\textsc{Saito Y., Itoh T.}:
\newblock {MusiCube}: {A} visual music recommendation system featuring interactive evolutionary computing.
\newblock In \emph{Proceedings of the 2011 Visual Information Communication-International Symposium} (2011), pp.~1--6.
\newblock \href {https://doi.org/10.1145/2016656.2016661} {\path{doi:10.1145/2016656.2016661}}.

\bibitem[SJL{\etalchar{*}}18]{stein2018}
\textsc{Stein M., Janetzko H., Lamprecht A., Breitkreutz T., Zimmermann P., Goldlücke B., Schreck T., Andrienko G., Grossniklaus M., Keim D.~A.}:
\newblock Bring it to the pitch: Combining video and movement data to enhance team sport analysis.
\newblock \emph{IEEE Transactions on Visualization and Computer Graphics 24}, 1 (2018), 13--22.
\newblock \href {https://doi.org/10.1109/TVCG.2017.2745181} {\path{doi:10.1109/TVCG.2017.2745181}}.

\bibitem[SMM12]{sedlmair2012}
\textsc{Sedlmair M., Meyer M., Munzner T.}:
\newblock Design study methodology: Reflections from the trenches and the stacks.
\newblock \emph{IEEE Transactions on Visualization and Computer Graphics 18}, 12 (2012), 2431--2440.
\newblock \href {https://doi.org/10.1109/TVCG.2012.213} {\path{doi:10.1109/TVCG.2012.213}}.

\bibitem[SMV13]{schulz2013}
\textsc{Schulz A., Matusik W., Velho L.}:
\newblock Choreographics: An authoring tool for dance shows.
\newblock \emph{Journal of Graphics Tools 17}, 4 (2013), 159--176.
\newblock \href {https://doi.org/10.1080/2165347X.2014.909341} {\path{doi:10.1080/2165347X.2014.909341}}.

\bibitem[{Sta}23]{stagekeep}
\textsc{{StageKeep}}:
\newblock {StageKeep}: Collaboration for choreographers.
\newblock \url{https://stagekeep.com/}, 2023.
\newblock Last accessed on 2023-11-30.

\bibitem[Ste23]{arrangeus}
\textsc{Stefanchuk A.}:
\newblock {ArrangeUs}: Dance formations made easy.
\newblock \url{https://apps.apple.com/de/app/arrangeus/id1502182540}, 2023.
\newblock Last accessed on 2023-11-30.

\bibitem[TLW{\etalchar{*}}20]{tang2020plotthread}
\textsc{Tang T., Li R., Wu X., Liu S., Knittel J., Koch S., Ertl T., Yu L., Ren P., Wu Y.}:
\newblock Plotthread: Creating expressive storyline visualizations using reinforcement learning.
\newblock \emph{IEEE Transactions on Visualization and Computer Graphics 27}, 2 (2020), 294--303.
\newblock \href {https://doi.org/10.1109/TVCG.2020.3030467} {\path{doi:10.1109/TVCG.2020.3030467}}.

\bibitem[TSK{\etalchar{*}}15]{tanisaro2015visual}
\textsc{Tanisaro P., Sch{\"o}ning J., Kurzhals K., Heidemann G., Weiskopf D.}:
\newblock Visual analytics for video applications.
\newblock \emph{it-Information Technology 57}, 1 (2015), 30--36.
\newblock \href {https://doi.org/10.1515/itit-2014-1072} {\path{doi:10.1515/itit-2014-1072}}.

\bibitem[WLS{\etalchar{*}}17]{wu2017ittvis}
\textsc{Wu Y., Lan J., Shu X., Ji C., Zhao K., Wang J., Zhang H.}:
\newblock {iTTVis}: {I}nteractive visualization of table tennis data.
\newblock \emph{IEEE Transactions on Visualization and Computer Graphics 24}, 1 (2017), 709--718.
\newblock \href {https://doi.org/10.1109/TVCG.2017.2744218} {\path{doi:10.1109/TVCG.2017.2744218}}.

\bibitem[Woo97]{wood1997semi}
\textsc{Wood L.~E.}:
\newblock Semi-structured interviewing for user-centered design.
\newblock \emph{interactions 4}, 2 (1997), 48--61.
\newblock \href {https://doi.org/10.1145/245129.245134} {\path{doi:10.1145/245129.245134}}.

\bibitem[Woo15]{wood2015visualizing}
\textsc{Wood J.}:
\newblock Visualizing personal progress in participatory sports cycling events.
\newblock \emph{IEEE Computer Graphics and Applications 35}, 4 (2015), 73--81.
\newblock \href {https://doi.org/10.1109/MCG.2015.71} {\path{doi:10.1109/MCG.2015.71}}.

\bibitem[{Wor}23]{WDSF2023}
\textsc{{World DanceSport Federation (WDSF)}}:
\newblock Wdsf competition rules.
\newblock \url{https://dancesport.app.box.com/s/j1thh09bgbg1ugcbr5zj}, 2023.

\end{thebibliography}


\end{document}